\documentclass[12pt,a4paper,english,amssymb,nofootinbib,superscriptaddress]{revtex4}
\usepackage{lmodern}
\usepackage{lmodern}

\usepackage[T1]{fontenc}
\usepackage[latin9]{inputenc}
\usepackage{babel}
\usepackage{mathrsfs}
\usepackage{tcolorbox}
\usepackage{amsmath}
\usepackage{amssymb}
\usepackage{graphicx}
\usepackage{esint}
\usepackage[unicode=true,pdfusetitle,
 bookmarks=true,bookmarksnumbered=false,bookmarksopen=false,
 breaklinks=false,pdfborder={0 0 1},backref=false,colorlinks=false]
 {hyperref}

\makeatletter

\@ifundefined{textcolor}{}
{%
 \definecolor{BLACK}{gray}{0}
 \definecolor{WHITE}{gray}{1}
 \definecolor{RED}{rgb}{1,0,0}
 \definecolor{GREEN}{rgb}{0,1,0}
 \definecolor{BLUE}{rgb}{0,0,1}
 \definecolor{CYAN}{cmyk}{1,0,0,0}
 \definecolor{MAGENTA}{cmyk}{0,1,0,0}
 \definecolor{YELLOW}{cmyk}{0,0,1,0}
}

\usepackage{babel}

\def\be{\begin{equation}}
\def\ee{\end{equation}}

\@ifundefined{textcolor}{}{%
	\definecolor{BLACK}{gray}{0}
	\definecolor{WHITE}{gray}{1}
	\definecolor{RED}{rgb}{1,0,0}
	\definecolor{GREEN}{rgb}{0,1,0}
	\definecolor{BLUE}{rgb}{0,0,1}
	\definecolor{CYAN}{cmyk}{1,0,0,0}
	\definecolor{MAGENTA}{cmyk}{0,1,0,0}
	\definecolor{YELLOW}{cmyk}{0,0,1,0}
}

\usepackage{latexsym}\usepackage{bm}

\makeatother

\begin{document}
\title{Non-stationary Energy of Perfect Fluid Sources in General Relativity}
\author{Emel Altas}
\email{emelaltas@kmu.edu.tr}

\affiliation{Department of Physics,\\
 Karamanoglu Mehmetbey University, 70100, Karaman, Turkey}
\author{Bayram Tekin}
\email{btekin@metu.edu.tr}

\affiliation{Department of Physics,\\
 Middle East Technical University, 06800, Ankara, Turkey}
\date{\today}
\begin{abstract}
\noindent The ADM energy for asymptotically flat spacetimes or its
generalizations to asymptotically non-flat spacetimes measure the
energy content of a stationary spacetime, such as a single black hole.
Such a stationary energy is given as a geometric invariant of the
spatial hypersurface of the spacetime and is expressed as an integral
on the boundary of the hypersurface. For non-stationary spacetimes,
there is a refinement of the ADM energy, the so-called Dain's invariant
that measures the non-stationary part, the gravitational radiation
component, of the total energy. Dain's invariant uses the metric and
the extrinsic curvature of the spatial hypersurface together with
the so-called approximate Killing initial data and vanishes for stationary
spacetimes. In our earlier work {[}Phys.Rev.D \textbf{{101}} (2020)
2, 024035{]}, we gave a reformulation of the non-stationary energy
for vacuum spacetimes in the Hamiltonian form of General Relativity
written succinctly in the Fischer-Marsden form. That formulation is
relevant for merging black holes or other compact sources. Here we
extend this formulation to non-vacuum spacetimes with a perfect fluid
source. This is expected to be relevant for spacetimes that have a
compact star, say a neutron star colliding with a black hole or another
non-vacuum object. 
\end{abstract}
\maketitle

\section{Introduction and a brief recapitulation of Dain's invariant in two different formulations}

The main purpose of this work is to derive an expression of non-stationary energy contained in a co-dimension one spacelike hypersurface in matter-coupled General Relativity, where the matter sector is taken to be a perfect fluid. {[}Our formulation will be valid for any type of source, but we shall give explicit results only in the perfect fluid case. {]}
As this discussion is a natural extension of the vacuum case, we will first recap what has been done so far in that case. Let us first note that some of what we shall briefly discuss here can also be found in our work \cite{Altas_S}, which we closely follow; but as the non-stationary energy concept and Dain's invariant \cite{Dain} are not widely known, it pays to summarize it here.

On a spacelike hypersurface $\Sigma$ of the spacetime, which we assume has the topology $\mathscr{M}=\mathbb{R}\times\Sigma$, one takes initial data to be the Riemannian metric $\gamma$ and the extrinsic curvature $K$ on $\Sigma$ for Einstein's gravity. We shall work in some local coordinates and so denote the components of the hypersurface metric as $\gamma_{ij}$ and the symmetric extrinsic curvature as
$K_{ij}$ with the indices taking values as $i,j=1,2,..,D-1$. Let
${\mathcal{D}}_{i}$ be the covariant derivative compatible with $\gamma_{ij}$;
and consider the usual lapse-shift decomposition of the metric as
(See figure 1) 
\begin{equation}
ds^{2}=\left(N_{i}N^{i}-N^{2}\right)dt^{2}+2N_{i}dtdx^{i}+\gamma_{ij}dx^{i}dx^{j}.\label{ADMdecompositionofmetric}
\end{equation}

\begin{figure}
~~\includegraphics[scale=0.5]{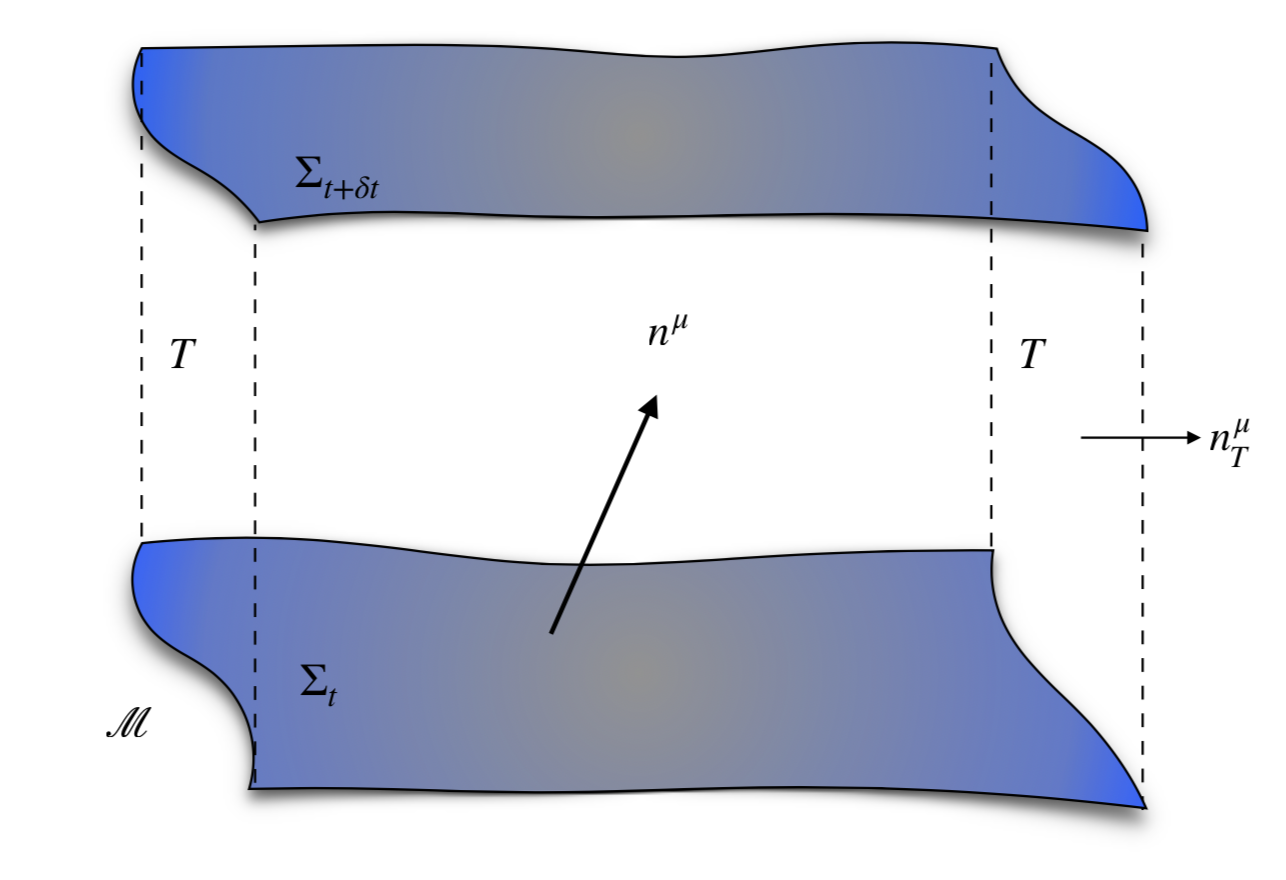}

\caption{Slicing of the spacetime in terms of co-dimension one spatial hypersurface.}
\end{figure}

\noindent In our conventions, the extrinsic curvature components read
explicitly as 
\begin{equation}
K_{ij}=\frac{1}{2N}\left(\dot{\gamma}_{ij}-{\mathcal{D}}_{i}N_{j}-{\mathcal{D}}_{j}N_{i}\right),\hskip1cm\dot{\gamma}_{ij}:=\partial_{t}\gamma_{ij}.
\end{equation}
With this spacetime decomposition, one can now almost forget about
the full covariant structure of spacetime and discuss everything in
terms of the tensor fields living and evolving on the hypersurface.
To this end, one can raise and lower the indices with the spatial
metric and its inverse. In particular, the trace of the extrinsic
curvature is defined as $K:=\gamma^{ij}K_{ij}$. For completeness,
and not to disturb the flow of the paper, we give a rather comprehensive
discussion of the ADM formulation \cite{ADM} in the Appendices.

Under the above decomposition of spacetime of which the details are
given in the Appendices, the $D$-dimensional Einstein equations with
a cosmological constant and a source term 
\begin{equation}
R_{\mu\nu}-\frac{1}{2}Rg_{\mu\nu}+\Lambda g_{\mu\nu}=\kappa T_{\mu\nu},
\end{equation}
produce the following $D$ constraints (the Hamiltonian and the momenta
constraints) on the hypersurface: 
\begin{eqnarray}
 &  & \Phi_{0}(\gamma,K):=-{}^{\Sigma}R-K^{2}+K^{ij}K_{ij}+2\Lambda-2\kappa T_{nn}=0,\nonumber \\
 &  & \Phi_{i}(\gamma,K):=-2{\mathcal{D}}_{k}K_{i}^{k}+2{\mathcal{D}}_{i}K-2\kappa T_{ni}=0,~~~~~~~~~~~~~~\label{Einstein_c}
\end{eqnarray}
where $^{\Sigma}R$ denotes the scalar curvature of the hypersurface;
and the energy-momentum tensor of the matter sector has the following
projections: 
\begin{eqnarray}
T_{nn} & = & \frac{1}{N^{2}}\left(2N^{i}T_{0i}-T_{00}-N^{i}N^{j}T_{ij}\right),\nonumber \\
T_{ni} & = & \frac{1}{N}\left(N^{j}T_{ij}-T_{0i}\right).
\end{eqnarray}
In addition to these constraints, we have the time evolution equations
which we shall give below.

\subsubsection{Dain's invariant using the constraints}

Let us briefly summarize Dain's original construction \cite{Dain}
of the non-stationary energy component of the total energy contained
in the initial data surface for the vacuum case. Hence one sets $T_{\mu\nu}=0$,
but we keep the cosmological constant slightly extending Dain's result,
and we do the computation in generic $D$ dimensions, extending the
four-dimensional result of \cite{Dain}.

Let the constraint covector be ${\bf \Phi}(\gamma,K):=(\Phi_{0},\Phi_{i})$,
and let ${\bf D}{\bf \Phi}(\gamma,K)$ be its linearization about
a given solution $(\gamma,K)$ to the constraints and ${\bf D\Phi}^{*}(\gamma,K)$
be the formal adjoint. Dain used an operator defined by Bartnik \cite{Bartnik}.
 Bartnik's operator looks somewhat mysterious at first sight, but later in the computation one realizes that it naturally should appear, and in fact, it is necessary to use. Without further ado, let us define
it
\begin{equation}
{\cal {P}}:={\bf D\Phi}(\gamma,K)\circ\begin{pmatrix}1 & 0\\
0 & -{\mathcal{D}}^{m},
\end{pmatrix},\hskip1cm\text{Bartnik's operator.}
\end{equation}
We also need the formal adjoint ${\cal {P}}^{*}$ of this operator to define the Dain's invariant on the codimension one hypersurface
as 
\begin{tcolorbox}
\begin{equation}
\mathscr{I}(\xi):=\intop_{\Sigma}dV~{\cal {P}}^{*}(\xi)\cdot{\cal {P}}^{*}(\xi),\hskip1cm\text{Dain's invariant}.\label{integralinvariant}
\end{equation}
\end{tcolorbox}
 Here $\xi:=(N,N^{i})$, thus ${P}^{*}(\xi):={P}^{*}\begin{pmatrix}N\\
N^{k}
\end{pmatrix}$ and the multiplication in (\ref{integralinvariant}) is defined component-wise
as follows 
\begin{equation}
\begin{pmatrix}N\\
N^{i}
\end{pmatrix}\cdot\begin{pmatrix}A\\
B_{i}
\end{pmatrix}:=NA+N^{i}B_{i}.
\end{equation}
The important point here is the following: the integral (\ref{integralinvariant})
is not to be computed for arbitrary lapse and shift $(N,N^{i})$ functions,
but for specific vectors $\xi:=(N,N^{i})$ that satisfy the following
fourth-order PDE 
\begin{tcolorbox}
\begin{equation}
{\cal {P}}\circ{\cal {P}}^{*}\left(\xi\right)=0,\hskip1cm\text{approximate KID equation.}\label{approx_KID}
\end{equation}
\end{tcolorbox}
 Dain dubbed this last equation as the "approximate Killing initial data" (KID) equation and in the case of time-symmetric initial data
$(K_{ij}=0)$, he showed that for any asymptotically flat three-manifold,
 the approximate KID equation has non-trivial solutions, that are solutions which only solve the full fourth-order equation. Of course, one must be careful here in stating what he proved: note that if $\xi$ satisfies the second order equation ${\cal {P}}^{*}\left(\xi\right)=0$, then it also automatically satisfies (\ref{approx_KID}). But, after some
reasonable decay assumptions at infinity, this second-order equation
can be shown to be the same as the first-order KID equation: ${\bf D\Phi}^{*}(\gamma,K)\left(\xi\right)=0$.
An important result about this is due to Moncrief \cite{Moncrief},
who proved that $\xi$ is a spacetime Killing vector satisfying $\mathcal{L}_{\xi}g=0$,
that is $\xi$ generates infinitesimal isometries \textit{if only
if} it satisfies the KID equations: 
\begin{tcolorbox}
\begin{equation}
\nabla_{\mu}\xi_{\nu}+\nabla_{\nu}\xi_{\mu}=0\Longleftrightarrow{\bf D\Phi^{*}}(\gamma,K)\left(\xi\right)=0,\hskip0.5cm\text{KIDs are Killing vectors.}
\end{equation}
\end{tcolorbox}
 From the physics vantage point Moncrief's theorem heuristically says
that, as expected, the isometries of the full spacetime are certainly
encoded in the initial data; and in this construction, the components
of the Killing vector are given simply by the lapse and the shift
functions.

\noindent From the above discussion, it is clear that Dain's invariant
(\ref{integralinvariant}), by construction, vanishes identically
when $\xi$ is a Killing vector, i.e. ${\mathcal{P}}\left(\xi\right)=0$
and the spacetime has exact symmetries. On the other hand, for approximate
translational KIDs, Dain argued that for asymptotically flat spaces,
and time-symmetric initial data, $\mathscr{I}(\xi)$ is a measure
of the non-stationary energy contained in the hypersurface $\Sigma$.
This non-stationary component is expected to evolve into gravitational
radiation in spacetime. The extension to time non-symmetric initial
data was carried out by Kroon and Williams \cite{Kroon}, where several
important results on KIDs by Moncrief \cite{Moncrief} and Beig-Chru\'{s}ciel
\cite{Beig} were used. Another formulation of the non-stationary
energy employing the time-evolution equations was given in \cite{Altas_S}
in generic $D$ dimensions and for spacetimes that are not necessarily
asymptotically flat. In the latter formulation, which makes use of
the approximate KIDs and the Hamiltonian formulation of General Relativity
in the compact Fischer-Marsden form, the physical meaning of the invariant
is more transparent. In both formulations, one can reduce the integral
to a co-dimension two spatial hypersurface after integration by parts.
The final formula is somewhat cumbersome, the reader is referred to
equation (53) of the work \cite{Altas_S} for the final result and
its various subcases. More recently, explicit details and extension
of Dain's invariant to the initial data describing black holes were
carried out by Sansom and Kroon \cite{Kroon3}.

As we shall need the basics of the formulation of Dain's invariant
using the time-evolution equations, let us briefly summarize the relevant
discussion given in \cite{Altas_S} here.

\subsubsection{Non-Stationary Energy via Time-evolution Equations}

Let the canonical phase space fields be the spatial metric $\gamma_{ij}$
and the canonical momenta $\pi^{ij}$. The Einstein-Hilbert Lagrangian
in the ADM formulation up to a boundary term reads 
\begin{equation}
\text{\ensuremath{\mathscr{L}}}_{EH}=\frac{1}{2\kappa}\sqrt{-g}\left(R-2\Lambda\right)=\frac{1}{2\kappa}\sqrt{\gamma}N\left(^{\Sigma}R+K_{ij}K^{ij}-K^{2}-2\Lambda\right)+\text{boundary~terms}.
\end{equation}
Then, by definition, one has 
\begin{equation}
\pi^{ij}:=\frac{\delta\text{\ensuremath{\mathscr{L}}}_{EH}}{\delta\dot{\gamma_{ij}}}=\frac{1}{2\kappa}\sqrt{\gamma}\left(K^{ij}-\gamma^{ij}K\right),\hskip1cm\pi=\frac{1}{2\kappa}\sqrt{\gamma}\left(2-D\right)K,
\end{equation}
with the reverse relations, for $D\ne2$, given as 
\begin{equation}
K^{ij}=\frac{2\kappa}{\sqrt{\gamma}}\left(\pi^{ij}-\frac{1}{D-2}\gamma^{ij}\pi\right),\hskip1cmK=\frac{2\kappa}{\sqrt{\gamma}\left(2-D\right)}\pi.
\end{equation}
The densitized version of the Hamiltonian and the momenta constraints
(\ref{Einstein_c}) for the case of pure gravity (no matter fields)
in terms of the canonical fields become 
\begin{eqnarray}
 &  & \Phi_{0}(\gamma,\pi):=\frac{\sqrt{\gamma}}{2\kappa}\left(-^{\Sigma}R+2\Lambda\right)+\frac{2\kappa}{\sqrt{\gamma}}\left(\pi_{ij}\pi^{ij}-\frac{\pi^{2}}{D-2}\right)=0,\nonumber \\
 &  & \Phi_{i}(\gamma,\pi):=-2\gamma_{ik}{\mathcal{D}}_{j}\pi^{kj}=0.\label{einstein_c2}
\end{eqnarray}
As explained in detail in \cite{Altas_S}, the Hamiltonian form of
the Einstein-Hilbert action, when extremized, leads to the Fischer-Marsden
form \cite{Fischer-Marsden} of the field equations 
\begin{tcolorbox}
\begin{equation}
\frac{d}{dt}\begin{pmatrix}\gamma\\
\pi
\end{pmatrix}=J\circ{\bf D\Phi^{*}}(\gamma,\pi)({\cal {N}}),\hskip1cmJ:=\begin{pmatrix}0 & 1\\
-1 & 0
\end{pmatrix}.\label{evolution}
\end{equation}
\end{tcolorbox}
 Here ${\cal {N}}$ is the lapse-shift vector with components $(N,N^{i})$.
A crucial point here is the following: the formal adjoint of the linearized
constraint map ${\bf D\Phi^{*}}(\gamma,\pi)$ appears in the time
evolution instead of the operator itself, and the constraints not
only determine the initial data, they also determine the time evolution.
The symplectic structure of the Hamiltonian equations is also evident
from the $J$ matrix. The constraints (\ref{einstein_c2}) augmented
with tensor equations (\ref{evolution}) constitute constrained dynamical
systems for a given lapse-shift vector $(N,N^{i})$. This form of
the equations is the most suitable one for our purpose since, as discussed
above, if ${\bf D\Phi^{*}}(\gamma,\pi)({\cal {N}})=0$, that is ${\cal {N}}=\xi$
is a Killing vector, then the time evolution is trivial. On the other
hand, if the lapse-shift vector is not a Killing vector, then one
can ask how much it fails to be a Killing vector by the following
equation 
\begin{equation}
{\bf D\Phi^{*}}(\gamma,\pi)({\cal {N}})=J^{-1}\circ\frac{d}{dt}\begin{pmatrix}\gamma\\
\pi
\end{pmatrix}.
\end{equation}
In particular, one can try to understand that the approximate KID
equation as defined by Dain in terms of the data on the hypersurface
can now have a different representation. For this purpose, we still
need to work a little more. For example, to match the dimensions,
and to get a number out of the above matrix, it was argued in \cite{Altas_S}
that one necessarily introduces the adjoint of the Bartnik's operator
\cite{Bartnik}: 
\begin{equation}
{\cal {P}}^{*}({\cal {N}}):=\begin{pmatrix}1 & 0\\
0 & {\mathcal{D}}_{m}
\end{pmatrix}\circ{\bf D\Phi^{*}}(\gamma,\pi)({\cal {N}})=\begin{pmatrix}1 & 0\\
0 & {\mathcal{D}}_{m}
\end{pmatrix}\circ J^{-1}\circ\frac{d}{dt}\begin{pmatrix}\gamma\\
\pi
\end{pmatrix},
\end{equation}
which at the end boils down to a very simple form ${\cal {P}}^{*}({\cal {N}})=(-\dot{\pi},{\mathcal{D}}_{m}\dot{\gamma})$.
But this is still not sufficient yet, as $\pi$ is a tensor density,
we define 
\begin{equation}
\widetilde{{\cal {P}}}^{*}({\cal {N}}):=\begin{pmatrix}\gamma^{-1/2} & 0\\
0 & 1
\end{pmatrix}\circ{\cal {P}}^{*}({\cal {N}}).\label{formaladjoint_ptilde}
\end{equation}
Finally we have another representation of Dain's invariant (actually
its generalization) which makes use of the time derivatives of the
canonical phase space variables: 
\begin{tcolorbox}
\begin{equation}
\text{\ensuremath{\mathscr{I}}}({\cal {N}})=\intop_{\Sigma}dV\thinspace\widetilde{{\cal {P}}}^{*}({\cal {N}})\cdot\widetilde{{\cal {P}}}^{*}({\cal {N}})=\intop_{\Sigma}dV\thinspace\left(|{\mathcal{D}}_{m}\dot{\gamma}_{ij}|^{2}+\frac{1}{\gamma}|\dot{\pi}^{ij}|^{2}\right),\label{bizimdenk}
\end{equation}
\end{tcolorbox}
\noindent where we have used the short-hand notations for squares as $|{\mathcal{D}}_{m}\dot{\gamma}_{ij}|^{2}:=\gamma^{mn}{\gamma}^{ij}{\gamma}^{kl}{\mathcal{D}}_{m}\dot{\gamma}_{ik}{\mathcal{D}}_{n}\dot{\gamma}_{jl}$
and $|\dot{\pi}^{ij}|^{2}:={\gamma}_{ij}{\gamma}_{kl}\dot{\pi}^{ik}\dot{\pi}^{jl}$.
Several remarks are apt here: the above integral is valid for any
lapse-shift vector and in the presence of a cosmological constant.
Observe that the time derivative of the canonical momentum appears
in the integral as well as the time derivative of the spatial covariant
derivative of the spatial metric, both of which vanish for the stationary
case. The integrand is explicitly positive definite. Moreover, when
$({\cal {N}})$ is an approximate KID, then something special happens
and one can turn this volume integral into a surface integral reproducing
the case of Dain. For details on this see \cite{Altas_S}.

\section{TIME EVOLUTION EQUATIONS, INCLUSION OF MATTER}

\noindent To be able to extend the discussion of non-stationary energy
to the non-vacuum case, here, we first find the time evolution equations
directly, without using the linearized constraint map. Starting from
the definition of the extrinsic curvature, time evolution of the dynamical
variable $\gamma_{ij}$, the spatial metric, reads 
\begin{equation}
\frac{d\gamma_{ij}}{dt}=2NK_{ij}+2\mathcal{D}_{(i}N_{j)},\label{gammadot}
\end{equation}
where we use the symmetrization notation with a 1/2 factor. Equivalently,
in terms of the conjugate momenta, one has 
\begin{equation}
\frac{d\gamma_{ij}}{dt}=4\kappa N{\mathcal{G}}_{ijkl}\,\pi^{kl}+2\mathcal{D}_{(i}N_{j)},
\end{equation}
where the \textit{DeWitt metric} \cite{DeWitt} ${\mathcal{G}}_{ijkl}$
in $D$ dimensions reads 
\begin{equation}
{\mathcal{G}}_{ijkl}=\frac{1}{2\sqrt{\gamma}}\left(\gamma_{ik}\gamma_{jl}+\gamma_{il}\gamma_{jk}-\frac{2}{D-2}\gamma_{ij}\gamma_{kl}\right).
\end{equation}
To find the evolution of the conjugate momentum, we consider the purely
spatial components of the cosmological Einstein equations 
\begin{equation}
R_{ij}-\frac{1}{2}R\gamma_{ij}+\Lambda\gamma_{ij}=\kappa T_{ij}.\label{fieldeqspatial}
\end{equation}
Firstly, inserting the ADM decomposition of the corresponding tensor
fields 
\begin{equation}
R_{ij}={}^{\Sigma}R_{ij}+KK_{ij}-2K_{ik}K_{j}^{k}+\frac{1}{N}\left(\dot{K}_{ij}-N^{k}\mathcal{D}_{k}K_{ij}-\mathcal{D}_{i}\mathcal{D}_{j}N-2K_{k(i}\mathcal{D}_{j)}N^{k}\right),\label{ricciij}
\end{equation}
and 
\begin{equation}
R=^{\Sigma}R+K^{2}+K_{ij}K^{ij}+\frac{2}{N}\left(\dot{K}-\mathcal{D}_{k}\mathcal{D}^{k}N-N^{k}\mathcal{D}_{k}K\right)\label{r-1}
\end{equation}
in (\ref{fieldeqspatial}), and then using the Hamiltonian constraint,
one arrives at 
\begin{eqnarray}
 &  & \dot{K}_{ij}-\gamma_{ij}\dot{K}=N(-{}^{\Sigma}R_{ij}-KK_{ij}+2K_{ik}K_{j}^{k}+\gamma_{ij}K_{kl}^{2})+N^{k}\mathcal{D}_{k}K_{ij}+\mathcal{D}_{i}D_{j}N\\
 &  & +2K_{k(i}\mathcal{D}_{j)}N^{k}-\gamma_{ij}\left(\mathcal{D}_{k}\mathcal{D}^{k}N+N^{k}\mathcal{D}_{k}K\right)+\kappa NT_{ij}-\frac{\kappa}{N}\gamma_{ij}\left(2N^{k}T_{ok}-T_{00}-N^{l}N^{k}T_{lk}\right).\nonumber 
\end{eqnarray}
Adding $-\dot{\gamma}_{ij}K$ to both sides of the last equation,
one has 
\begin{eqnarray}
 &  & \frac{d}{dt}(K_{ij}-\gamma_{ij}K)=N\left(-^{\Sigma}R_{ij}-3KK_{ij}+2K_{ik}K_{j}^{k}+\gamma_{ij}K_{kl}^{2}\right)+N^{k}\mathcal{D}_{k}K_{ij}\nonumber \\
 &  & ~~~~~~~~~~~~~~~~~~~~~+\mathcal{D}_{i}\mathcal{D}_{j}N-2K\mathcal{D}_{(i}N_{j)}+2K_{k(i}\mathcal{D}_{j)}N^{k}-\gamma_{ij}\left(\mathcal{D}_{k}\mathcal{D}^{k}N+N^{k}\mathcal{D}_{k}K\right)\nonumber \\
 &  & ~~~~~~~~~~~~~~~~~~~~~+\kappa NT_{ij}-\frac{\kappa}{N}\gamma_{ij}\left(2N^{k}T_{ok}-T_{00}-N^{l}N^{k}T_{lk}\right).\label{eqN}
\end{eqnarray}
To obtain $\dot{\pi}_{ij}$, we multiply the result by $\sqrt{\gamma}$
and use 
\begin{equation}
\sqrt{\gamma}\frac{d}{dt}(K_{ij}-\gamma_{ij}K)=2\kappa\dot{\pi}_{ij}-\sqrt{\gamma}\left(NKK_{ij}+K_{ij}\mathcal{D}_{k}N^{k}-\gamma_{ij}NK^{2}-\gamma_{ij}K\mathcal{D}_{k}N^{k}\right).
\end{equation}
Then, equation (\ref{eqN}) can be rewritten in terms of conjugate
momentum as 
\begin{eqnarray}
 &  & \frac{d\pi_{ij}}{dt}=\frac{\sqrt{\gamma}}{2\kappa}\left(-N^{\Sigma}R_{ij}+\mathcal{D}_{i}\mathcal{D}_{j}N-\gamma_{ij}\mathcal{D}_{k}\mathcal{D}^{k}N\right)+\mathcal{D}_{k}(\pi_{ij}N^{k})+2\pi_{k(i}\mathcal{D}_{j)}N^{k}\nonumber \\
 &  & ~~~~~~~~~~+N\frac{2\kappa}{\sqrt{\gamma}}\left(2(\pi_{ik}\pi_{j}^{k}-\frac{\pi\pi_{ij}}{D-2})+\gamma_{ij}(\pi_{kl}^{2}-\frac{\pi^{2}}{D-2})\right)\nonumber \\
 &  & ~~~~~~~~~~+\frac{\sqrt{\gamma}}{2N}\left(N^{2}T_{ij}-\gamma_{ij}(2N^{k}T_{ok}-T_{00}-N^{l}N^{k}T_{lk})\right).
\end{eqnarray}
We will also need the up-up indices version of this. In terms of the
DeWitt metric, it reads 
\begin{tcolorbox}
\begin{eqnarray}
\frac{d\pi^{ij}}{dt}= &  & \frac{\sqrt{\gamma}}{2\kappa}\left(-N^{\Sigma}R^{ij}+\mathcal{D}^{i}\mathcal{D}^{j}N-\gamma^{ij}\mathcal{D}_{k}\mathcal{D}^{k}N\right)+\mathcal{D}_{k}(\pi^{ij}N^{k})-2\pi_{k}\thinspace^{(i}\mathcal{D}_{k}N^{j)}\nonumber \\
 &  & +N\frac{2\kappa}{\sqrt{\gamma}}\left(-2{\mathcal{G}}_{klmn}\gamma^{ik}\pi^{jl}\pi^{mn}+\gamma^{ij}{\mathcal{G}}_{klmn}\pi^{kl}\pi^{mn}\right)\nonumber \\
 &  & +\frac{\sqrt{\gamma}}{2N}\left(N^{2}T_{ij}-\gamma_{ij}(2N^{k}T_{ok}-T_{00}-N^{l}N^{k}T_{lk})\right).
\end{eqnarray}
\end{tcolorbox}
 Using the Hamiltonian constraint one more time, one can express the
last equation as 
\begin{eqnarray}
 &  & \frac{d\pi^{ij}}{dt}=\frac{\sqrt{\gamma}}{2\kappa}\left(-N^{\Sigma}\mathcal{G}^{ij}+\mathcal{D}^{i}\mathcal{D}^{j}N-\gamma^{ij}\mathcal{D}_{k}\mathcal{D}^{k}N\right)+\mathcal{L}_{N}\pi^{ij}+\pi^{ij}\mathcal{D}_{k}N^{k}\nonumber \\
 &  & ~~~~~~~~~~+N\frac{2\kappa}{\sqrt{\gamma}}\left(-2(\pi^{ik}\pi^{j}\thinspace_{k}-\frac{\pi\pi^{ij}}{D-2})+\frac{1}{2}\gamma^{ij}(\pi_{kl}^{2}-\frac{\pi^{2}}{D-2})\right)+\sqrt{\gamma}\frac{N}{2}T^{ij},
\end{eqnarray}
where we have used the hypersurface Einstein tensor given as $^{\Sigma}\mathcal{G}^{ij}={}^{\Sigma}R^{ij}-\frac{1}{2}\gamma^{ij}{}^{\Sigma}R+\Lambda\gamma^{ij}$;
and we also used the Lie-derivative along the shift- vector, $\mathcal{L}_{N}\pi^{ij}$,
that reads explicitly as 
\begin{equation}
\mathcal{L}_{N}\pi^{ij}:=N^{k}\mathcal{D}_{k}\pi^{ij}-\pi^{ki}\mathcal{D}_{k}N^{j}-\pi^{kj}\mathcal{D}_{k}N^{i}.
\end{equation}

\section{Non-stationary energy of PERFECT FLUIDS}

As a concrete and a useful application, let us study the case of a
perfect fluid source with the energy-momentum tensor given as 
\begin{equation}
T_{\mu\nu}=(\rho+p)u_{\mu}u_{\nu}+p\,g_{\mu\nu},
\end{equation}
where $p$ is the pressure, $\rho$ is the energy density and $u^{\mu}$
is the $D$-velocity of the perfect fluid. Following \cite{Marsden},
we take 
\begin{equation}
(\rho+p)=n\,h,
\end{equation}
where $n$ denotes the baryon number and $h$ denotes the enthalpy.
The fluid velocity can be taken as 
\begin{equation}
u^{\mu}=hJ^{\mu},
\end{equation}
and can be decomposed according to the hypersurface as 
\begin{equation}
J^{\mu}=J^{\bot}n^{\mu}+J_{\Vert}^{\mu}.
\end{equation}
The hypersurface orthogonal part reads $J^{\bot}:=-J^{\sigma}n_{\sigma}=J_{\bot}$.
Contracting the last equation with $n_{\mu}$ and using $n_{\mu}n^{\mu}=-1$,
we obtain $n_{\mu}J_{\Vert}^{\mu}=0$. Since 
\begin{equation}
n_{\mu}=(-N,0),\hskip1cmn^{\mu}=(1/N,-N^{i}/N),
\end{equation}
we can write $J_{\Vert}^{0}=0$ and $J_{\Vert0}=N^{i}J_{\Vert i}$.
If we evaluate the zeroth component, we find $J^{0}=J^{\bot}/N$,
and the lower index case reads 
\begin{equation}
J_{0}=-NJ^{\bot}+N^{i}J_{\Vert i},
\end{equation}
which yields 
\begin{equation}
J_{i}=J_{\Vert i}.
\end{equation}
These results reduce the spatial component of the energy momentum
tensor to 
\begin{equation}
T_{ij}=(\rho+p)u_{i}u_{j}+pg_{ij}=nh^{3}J_{\Vert i}J_{\Vert j}+p\gamma_{ij}.
\end{equation}
Then, the time evolution equation for the conjugate momentum reads
\begin{eqnarray}
 &  & \frac{d\pi^{ij}}{dt}=\frac{\sqrt{\gamma}}{2\kappa}\left(-N^{\Sigma}{\mathcal{G}}^{ij}+\mathcal{D}^{i}\mathcal{D}^{j}N-\gamma^{ij}\mathcal{D}_{k}\mathcal{D}^{k}N\right)+\mathcal{L}_{N}\pi^{ij}+\pi^{ij}\mathcal{D}_{k}N^{k}\nonumber \\
 &  & +N\frac{2\kappa}{\sqrt{\gamma}}\left(-2(\pi^{ik}\pi^{j}\thinspace_{k}-\frac{\pi\pi^{ij}}{D-2})+\frac{1}{2}\gamma^{ij}(\pi_{kl}^{2}-\frac{\pi^{2}}{D-2})\right)+\sqrt{\gamma}\frac{N}{2}\left(nh^{3}J_{\Vert}^{i}J_{\Vert}^{j}+p\gamma^{ij}\right).
\end{eqnarray}
Inserting these results in our formulation of the non-stationary energy
formula (\ref{bizimdenk}) 
\begin{equation}
\text{\ensuremath{\mathscr{I}}}({\cal {N}})=\intop_{\Sigma}dV\thinspace\left(|\mathcal{D}_{m}\dot{\gamma}_{ij}|^{2}+\frac{1}{\gamma}|\dot{\pi}^{ij}|^{2}\right),
\end{equation}
we have 
\begin{align}
|\mathcal{D}_{m}\dot{\gamma}_{ij}|^{2} & =\frac{16N^{2}\kappa^{2}}{\gamma}(\mathcal{D}_{m}\pi_{ij}\mathcal{D}^{m}\pi^{ij}+\frac{D-3}{D-2}\partial_{m}\pi\partial^{m}\pi)+4\mathcal{D}_{m}\mathcal{D}_{(i}N_{j)}\mathcal{D}^{m}\mathcal{D}^{(i}N^{j)}\nonumber \\
 & +\frac{16N\kappa}{\sqrt{\gamma}}\left(\mathcal{D}_{m}\pi_{ij}-\frac{\gamma_{ij}}{D-2}\partial_{m}\pi\right)\mathcal{D}^{m}\mathcal{D}^{(i}N^{j)}.
\end{align}
To recast the equation in a more compact form, let us define 
\begin{eqnarray}
 &  & X^{ij}:=\frac{\sqrt{\gamma}}{2\kappa}\left(-N^{\Sigma}\mathcal{G}^{ij}+\mathcal{D}^{i}\mathcal{D}^{j}N-\gamma^{ij}\mathcal{D}_{k}\mathcal{D}^{k}N\right),\nonumber \\
 &  & Y^{ij}:=N\frac{2\kappa}{\sqrt{\gamma}}\left(-2(\pi^{ik}\pi^{j}\thinspace_{k}-\frac{\pi\pi^{ij}}{D-2})+\frac{1}{2}\gamma^{ij}(\pi_{kl}^{2}-\frac{\pi^{2}}{D-2})\right),
\end{eqnarray}
so that we have 
\begin{equation}
\frac{d\pi^{ij}}{dt}=X^{ij}+\mathcal{L}_{N}\pi^{ij}+\pi^{ij}\mathcal{D}_{k}N^{k}+Y^{ij}+\sqrt{\gamma}\frac{N}{2}T^{ij},
\end{equation}
which yields 
\begin{align}
|\dot{\pi}^{ij}|^{2} & =X_{ij}^{2}+\gamma_{ik}\gamma_{jl}\mathcal{L}_{N}\pi^{kl}\mathcal{L}_{N}\pi^{ij}+\pi_{ij}^{2}\mathcal{D}_{k}N^{k}\mathcal{D}_{l}N^{l}+Y_{ij}^{2}+\gamma\frac{N^{2}}{4}T^{ij}T_{ij}\nonumber \\
 & +2X_{ij}\left(\mathcal{L}_{N}\pi^{ij}+\pi^{ij}\mathcal{D}_{k}N^{k}+Y^{ij}+\sqrt{\gamma}\frac{N}{2}T^{ij}\right)+2\mathcal{L}_{N}\pi^{ij}\left(\pi_{ij}\mathcal{D}_{k}N^{k}+Y_{ij}+\sqrt{\gamma}\frac{N}{2}T_{ij}\right)\nonumber \\
 & +2\pi^{ij}\mathcal{D}_{k}N^{k}\left(Y_{ij}+\sqrt{\gamma}\frac{N}{2}T_{ij}\right)+\sqrt{\gamma}NY_{ij}T^{ij},
\end{align}
where $X_{ij}^{2}=X_{ij}X^{ij}$. Collecting all the pieces, one arrives
then 
\begin{tcolorbox}
\begin{align}
\text{\ensuremath{\mathscr{I}}}({\cal {\xi}}) & =\intop_{\Sigma}dV\thinspace\Biggl(\frac{16N^{2}\kappa^{2}}{\gamma}(\mathcal{D}_{m}\pi_{ij}\mathcal{D}^{m}\pi^{ij}+\frac{D-3}{D-2}\partial_{m}\pi\partial^{m}\pi)+4\mathcal{D}_{m}\mathcal{D}_{(i}N_{j)}\mathcal{D}^{m}\mathcal{D}^{(i}N^{j)}\nonumber \\
 & +\frac{16N\kappa}{\sqrt{\gamma}}\left(\mathcal{D}_{m}\pi_{ij}-\frac{\gamma_{ij}}{D-2}\partial_{m}\pi\right)\mathcal{D}^{m}\mathcal{D}^{(i}N^{j)}\nonumber \\
 & +\frac{1}{\gamma}\left(X_{ij}^{2}+\gamma_{ik}\gamma_{jl}\mathcal{L}_{N}\pi^{kl}\mathcal{L}_{N}\pi^{ij}+\pi_{ij}^{2}\mathcal{D}_{k}N^{k}\mathcal{D}_{l}N^{l}+Y_{ij}^{2}+\gamma\frac{N^{2}}{4}T_{ij}^{2}\right)\nonumber \\
 & +\frac{2X_{ij}}{\gamma}\left(\mathcal{L}_{N}\pi^{ij}+\pi^{ij}\mathcal{D}_{k}N^{k}+Y^{ij}+\sqrt{\gamma}\frac{N}{2}T^{ij}\right)\nonumber \\
 & +\frac{2}{\gamma}\mathcal{L}_{N}\pi^{ij}\left(\pi_{ij}\mathcal{D}_{k}N^{k}+Y_{ij}+\sqrt{\gamma}\frac{N}{2}T_{ij}\right)\nonumber \\
 & +\frac{2}{\gamma}\pi^{ij}\mathcal{D}_{k}N^{k}\left(Y_{ij}+\sqrt{\gamma}\frac{N}{2}T_{ij}\right)+\frac{N}{\sqrt{\gamma}}Y_{ij}T^{ij}\Biggr),\label{finalf}
\end{align}
\end{tcolorbox}
 where we took $\xi$ to be an approximate KID satisfying (\ref{approx_KID}).

\section{Conclusions}

Fischer-Marsden form of Einstein equations can be seen as the failure of initial data to possess an exact time translation symmetry. This simple observation led us earlier \cite{Altas_S} to give another representation of Dain's invariant \cite{Dain} which was originally given in terms of the constraints and the approximate Killing initial data. In this work, we extended our discussion to the non-vacuum case and specifically discussed the non-stationary energy that can be assigned to a spacetime with a perfect fluid source. Of course, as expected the final formula (\ref{finalf}) is rather cumbersome, and further progress requires evaluating this expression in a given (numerical)
solution. 

Finally, let us note that while we pursued and generalized Dain's approach to the non-stationary case based on the notion of approximate Killing initial data and the Fischer-Marsden form of Einstein equations, there are other approaches to gravitational radiation, the two most prominent ones being due to Newman-Penrose \cite{NP} and Penrose\cite{Pen} that is based on the conformal compactification of null infinity ${\mathcal{I}}^+$; and the Bondi-Metzner-Sachs (BMS)\cite{BMS1,BMS2} approach based on the asymptotic structure of future null infinity germane to outgoing radiation. There must be an intimate connection between these approaches and the one we presented here. Especially, the BMS approach, which also gave rise to much recent work \cite{Strominger} regarding asymptotic symmetries, gravitational memory, and soft charges, seems so close in spirit to the formalism outlined here. But these connections are subtle at this stage and more work is needed. \footnote{We would like to thank an astute referee who brought our attention to these issues.} 

\section{{\normalsize{}{}{}{}Appendices}}

\subsection{{\normalsize{}{}{}{}ADM Split of Einstein's Equations in $D$
Dimensions}}

As our computations depend on the space+time splitting of Einstein's 
equations and all the relevant tensors, we here give the relevant details. Using the $\left(D-1\right)+1$
dimensional decomposition of the metric (\ref{ADMdecompositionofmetric})
we have: 
\begin{equation}
g_{00}=-N^{2}+N_{i}N^{i},~\ ~~~g_{0i}=N_{i},~~~~\ g_{ij}=\gamma_{ij},
\end{equation}
and  the inverse metric as 
\begin{equation}
g^{00}=-\frac{1}{N^{2}},~~~~g^{0i}=\frac{1}{N^{2}}N^{i},~~~~g^{ij}=\gamma^{ij}-\frac{1}{N^{2}}N^{i}N^{j}.
\end{equation}
And the determinant of the metric reads 
\begin{equation}
\sqrt{-g}=N\sqrt{\gamma},
\end{equation}
where we have used $g=\det g_{\mu\nu}$ and also $\gamma=\det\gamma_{ij}$.

Let $\Gamma_{\nu\rho}^{\mu}$ denote the Christoffel symbol of the
$D$ dimensional spacetime 
\begin{equation}
\Gamma_{\nu\rho}^{\mu}=\frac{1}{2}g^{\mu\sigma}\left(\partial_{\nu}g_{\rho\sigma}+\partial_{\rho}g_{\nu\sigma}-\partial_{\sigma}g_{\nu\rho}\right)
\end{equation}
and let $^{\Sigma}\Gamma_{ij}^{k}$  the Christoffel symbol
of the $D-1$ dimensional hypersurface,  that is compatible with the
spatial metric $\gamma_{ij}$:
\begin{equation}
^{\Sigma}\Gamma_{ij}^{k}=\frac{1}{2}\gamma^{kp}\left(\partial_{i}\gamma_{jp}+\partial_{j}\gamma_{ip}-\partial_{p}\gamma_{ij}\right).\label{Christoffelofhypersurface}
\end{equation}
Then  one can show the following relations
\begin{eqnarray}
&&\Gamma_{00}^{0}=\frac{1}{N}\left(\dot{N}+N^{k}(\partial_{k}N+N^{i}K_{ik})\right), \\
&&\Gamma_{0i}^{0}=\frac{1}{N}\left(\partial_{i}N+N^{k}K_{ik}\right),~~~~\Gamma_{ij}^{0}=\frac{1}{N}K_{ij},~~~~\Gamma_{ij}^{k}=^{\Sigma}\Gamma_{ij}^{k}-\frac{N^{k}}{N}K_{ij}, \\
&&\Gamma_{0j}^{i}=-\frac{1}{N}N^{i}\left(\partial_{j}N+K_{kj}N^{k}\right)+NK_{j}\thinspace^{i}+\mathcal{D}_{j}N^{i}, \\
&&\Gamma_{00}^{i}=-\frac{N^{i}}{N}\left(\dot{N}+N^{k}\left(\partial_{k}N+N^{l}K_{kl}\right)\right)+N\left(\partial^{i}N+2N^{k}K_{k}\thinspace^{i}\right)+\dot{N}^{i}+N^{k}\mathcal{D}_{k}N^{i}.
\end{eqnarray}
To compute the decomposition of the field equations, we need to express
additional tensor quantities such as the Ricci tensor components, the
scalar curvature.

\subsection{{\normalsize{}{}{}{}ADM split of the Ricci tensor and the scalar
curvature}}

Starting with the definition of the $D$ dimensional Ricci tensor
\begin{equation}
R_{\rho\sigma}=\partial_{\mu}\Gamma_{\rho\sigma}^{\mu}-\partial_{\rho}\Gamma_{\mu\sigma}^{\mu}+\Gamma_{\mu\nu}^{\mu}\Gamma_{\rho\sigma}^{\nu}-\Gamma_{\sigma\nu}^{\mu}\Gamma_{\mu\rho}^{\nu},
\end{equation}
one has 
\begin{eqnarray*}
 &  & R_{ij}=\partial_{0}\Gamma_{ij}^{0}+\partial_{k}\Gamma_{ij}^{k}-\partial_{i}(\Gamma_{0j}^{0}+\Gamma_{kj}^{k})+\Gamma_{ij}^{0}(\Gamma_{00}^{0}+\Gamma_{k0}^{k})\\
 &  & ~~~~~~+\Gamma_{ij}^{k}\Gamma_{0k}^{0}+\Gamma_{kl}^{k}\Gamma_{ij}^{l}-\Gamma_{0j}^{0}\Gamma_{0i}^{0}-\Gamma_{kj}^{0}\Gamma_{0i}^{k}-\Gamma_{ki}^{0}\Gamma_{0j}^{k}-\Gamma_{jl}^{k}\Gamma_{ki}^{l},
\end{eqnarray*}
which yields
\begin{equation}
R_{ij}={}^{\Sigma}R_{ij}+KK_{ij}-2K_{ik}K_{j}^{k}+\frac{1}{N}\left(\dot{K}_{ij}-N^{k}\mathcal{D}_{k}K_{ij}-\mathcal{D}_{i}\mathcal{D}_{j}N-K_{ki}\mathcal{D}_{j}N^{k}-K_{kj}\mathcal{D}_{i}N^{k}\right),\label{eq:rij}
\end{equation}
where $^{\Sigma}R_{ij}$ denotes the $ij$ component of the Ricci
tensor on the hypersurface  given as 
\begin{equation}
^{\Sigma}R_{ij}=\partial_{k}\thinspace^{\Sigma}\Gamma_{ij}^{k}-\partial_{i}\thinspace^{\Sigma}\Gamma_{kj}^{k}+\thinspace^{\Sigma}\Gamma_{kl}^{k}\thinspace^{\Sigma}\Gamma_{ij}^{l}-\thinspace^{\Sigma}\Gamma_{kl}^{k}\thinspace^{\Sigma}\Gamma_{ki}^{l}.
\end{equation}
The $0i$ component can be written as
\begin{equation}
R_{0i}=\partial_{0}\Gamma_{0i}^{0}+\partial_{k}\Gamma_{0i}^{k}-\partial_{i}(\Gamma_{00}^{0}+\Gamma_{k0}^{k})+\Gamma_{0i}^{0}\Gamma_{k0}^{k}+\Gamma_{kl}^{k}\Gamma_{i0}^{l}-\Gamma_{00}^{k}\Gamma_{ki}^{0}-\Gamma_{0l}^{k}\Gamma_{ki}^{l},
\end{equation}
and this expression gives us the following simple result 
\begin{equation}
R_{0i}=N^{j}R_{ij}+N\left(\mathcal{D}_{m}K_{i}^{m}-\mathcal{D}_{i}K\right).\label{eq:ri0}
\end{equation}
Similarly, the $00$ component 
\begin{equation}
R_{0i}=\partial_{k}\Gamma_{00}^{k}-\partial_{0}\Gamma_{0k}^{k}+\Gamma_{00}^{0}\Gamma_{k0}^{k}+\Gamma_{kl}^{k}\Gamma_{00}^{l}-\Gamma_{00}^{k}\Gamma_{k0}^{0}-\Gamma_{0l}^{k}\Gamma_{k0}^{l},
\end{equation}
can be written in a compact form as

\begin{equation}
R_{00}=N^{i}N^{j}R_{ij}-N^{2}K_{ij}K^{ij}+N\left(\mathcal{D}_{k}\mathcal{D}^{k}N-\dot{K}-N^{k}\mathcal{D}_{k}K+2N^{k}\mathcal{D}_{m}K_{k}^{m}\right).\label{r00}
\end{equation}
Then, the scalar curvature of the spacetime, $R=g^{\mu\nu}R_{\mu\nu}$,
can be expressed in terms of the scalar curvature of the spatial hypersurface,
$\thinspace^{\Sigma}R=\gamma^{ij}{}^{\Sigma}R_{ij}$, as

\begin{equation}
R=\thinspace^{\Sigma}R+K^{2}+K_{ij}K^{ij}+\frac{2}{N}\left(\dot{K}-\mathcal{D}_{k}\mathcal{D}^{k}N-N^{k}\mathcal{D}_{k}K\right).\label{r}
\end{equation}

\subsection{{\normalsize{}{}{}{}ADM Lagrangian density}}

The Einstein-Hilbert Lagrangian density reads
\begin{equation}
\text{\ensuremath{\mathscr{L}}}_{EH}=\frac{1}{2\kappa}\sqrt{-g}(R-2\Lambda).
\end{equation}
Inserting (\ref{r}), using the relation $\sqrt{-g}=N\sqrt{\gamma}$
together with
\begin{equation}
2\sqrt{\gamma}\dot{K}=\partial_{0}(2K\sqrt{\gamma})-\sqrt{\gamma}(2NK^{2}+2K\mathcal{D}_{k}N^{k}),
\end{equation}
one obtains the Lagrangian density as 
\begin{equation}
\text{\ensuremath{\mathscr{L}}}_{EH}=\frac{1}{\kappa}\left(\partial_{0}(K\sqrt{\gamma})-\mathcal{D}_{k}(\sqrt{\gamma}N^{k}K+\sqrt{\gamma}\partial^{k}N)\right)+\frac{1}{2\kappa}\sqrt{\gamma}N\left(^{\Sigma}R+K_{ij}^{2}-K^{2}-2\Lambda\right).
\end{equation}
Ignoring the boundary expression, we get 
\begin{equation}
\text{\ensuremath{\mathscr{L}}}_{EH}=\frac{1}{2\kappa}\sqrt{\gamma}N\left(^{\Sigma}R+K_{ij}^{2}-K^{2}-2\Lambda\right).
\end{equation}
The canonical momenta is, as usual, defined as follows 
\begin{equation}
\pi^{ij}:=\frac{\delta\text{\ensuremath{\mathscr{L}}}_{EH}}{\delta\dot{\gamma_{ij}}},
\end{equation}
and equivalently can be written as 
\begin{equation}
\pi^{ij}:=\frac{\delta\text{\ensuremath{\mathscr{L}}}_{EH}}{\delta\dot{\gamma_{ij}}}=\frac{\delta\text{\ensuremath{\mathscr{L}}}_{EH}}{\delta K_{kl}}\frac{\delta K_{kl}}{\delta\dot{\gamma_{ij}}},
\end{equation}
where the variation of the Lagrangian density yields
\begin{equation}
\frac{\delta\text{\ensuremath{\mathscr{L}}}_{EH}}{\delta K_{kl}}=\frac{N\sqrt{\gamma}}{\kappa}(K^{kl}-\gamma^{kl}K).
\end{equation}
Also, due to definition of the extrinsic curvature one obtains
\begin{equation}
\frac{\delta K_{kl}}{\delta\dot{\gamma_{ij}}}=\frac{1}{2N}\delta_{k}^{i}\delta_{l}^{j}.
\end{equation}
Collecting the pieces, one ends up with 
\begin{equation}
\pi^{ij}=\frac{1}{2\kappa}\sqrt{\gamma}(K^{ij}-\gamma^{ij}K).
\end{equation}
Taking the trace one has 
\begin{equation}
\pi=\frac{1}{2\kappa}\sqrt{\gamma}(2-D)K.
\end{equation}

\subsection{{\normalsize{}{}{}{}ADM Hamiltonian density}}

Einstein-Hilbert Hamiltonian density reads 
\begin{equation}
{\cal {H}}_{EH}=\pi^{ij}\dot{\gamma}_{ij}-\text{\ensuremath{\mathscr{L}}}_{EH}.
\end{equation}
Using the previous results it is straightforward to find it explicitly
\begin{align}
{\cal {H}}_{EH} & =\frac{1}{\kappa}\left(\mathcal{D}_{k}(\sqrt{\gamma}N^{i}K_{i}^{k}+\sqrt{\gamma}\partial^{k}N)-\partial_{0}(K\sqrt{\gamma})\right)\nonumber \\
 & +\frac{\sqrt{\gamma}N}{2\kappa}\left(-^{\Sigma}R+K_{ij}^{2}-K^{2}+2\Lambda\right)+\frac{\sqrt{\gamma}N^{i}}{2\kappa}(\mathcal{D}_{i}K-\mathcal{D}_{k}K_{i}^{k}).
\end{align}
Here the first three terms on the right hand side of the equality
are boundary terms and they do not contribute to the constraint equations.

\subsection{{\normalsize{}{}{}{}ADM Hamiltonian and constraint equations}}

Up to a boundary expression ADM Hamiltonian yields the constraints
\begin{equation}
H_{EH}=\intop_{V}dV \thinspace{\cal {H}}_{EH}=\intop_{V}dV\thinspace\left(N\thinspace\Phi_{0}+N^{i}\thinspace\Phi_{i}\right),
\end{equation}
where $\Phi_{0}$ denotes the Hamiltonian constraint and $\Phi_{i}$
denotes the momentum constraint. One explicitly gets
\begin{equation}
H_{EH}=\intop_{V}dV \left (\thinspace\frac{\sqrt{\gamma}N}{2\kappa}\left(-^{\Sigma}R+K_{ij}^{2}-K^{2}+2\Lambda\right)+\frac{\sqrt{\gamma}N^{i}}{2\kappa}(\mathcal{D}_{i}K-\mathcal{D}_{k}K_{i}^{k})\right),
\end{equation}
yielding the Hamiltonian constraint as
\begin{equation}
\Phi_{0}(\gamma,K)=\frac{\sqrt{\gamma}}{2\kappa}\left(-^{\Sigma}R+K_{ij}^{2}-K^{2}+2\Lambda\right),
\end{equation}
and also the momentum constraint as
\begin{equation}
\Phi_{i}(\gamma,K)=\frac{\sqrt{\gamma}}{2\kappa}(\mathcal{D}_{i}K-\mathcal{D}_{k}K_{i}^{k}).
\end{equation}
In terms of the conjugate momenta, using the reverse relations 
\begin{equation}
K^{ij}=\frac{2\kappa}{\sqrt{\gamma}}(\pi^{ij}-\gamma^{ij}\pi),\thinspace\thinspace\thinspace\thinspace\thinspace\thinspace\thinspace\thinspace\thinspace\thinspace\thinspace\thinspace\thinspace\thinspace\thinspace\thinspace\thinspace\thinspace K^{ij}=-\frac{2\kappa}{\sqrt{\gamma}(D-2)}\pi.
\end{equation}
we can equivalently write the following equations
\begin{equation}
\Phi_{0}(\gamma,\pi)=\frac{\sqrt{\gamma}}{2\kappa}\left(-^{\Sigma}R+2\Lambda\right)+\frac{2\kappa}{\sqrt{\gamma}}(\pi_{ij}^{2}-\frac{\pi^{2}}{D-2}),
\end{equation}
\begin{equation}
\Phi_{i}(\gamma,\pi)=-2\mathcal{D}_{k}\pi_{i}^{k}.
\end{equation}

\subsection{{\normalsize{}{}{}{}Constraint equations via field equations}}

We can also get the constraints directly from the field Einstein equations: 
\begin{equation}
R_{\mu\nu}-\frac{1}{2}g_{\mu\nu}R+\varLambda g_{\mu\nu}=\kappa T_{\mu\nu}
\end{equation}
which split into the constraints and the evolution equations.
Obviously we have 
\begin{equation}
R_{ij}-\frac{1}{2}\gamma_{ij}R+\varLambda\gamma_{ij}=\kappa T_{ij},\label{hypersurefaceij}
\end{equation}
which can be used to simplify the constraint equations. Starting from the $0i$ component, we have 
\begin{equation}
R_{0i}-\frac{1}{2}g_{0i}R+\varLambda g_{0i}=\kappa T_{0i}.
\end{equation}
Using $g_{0i}=N_{i}$ and plugging the ADM decomposition of the $0i$
component of the Ricci tensor one obtains 
\begin{equation}
N^{j}\left(R_{ij}-\frac{1}{2}\gamma_{ij}R+\varLambda\gamma_{ij}\right)+N(\mathcal{D}_{k}K_{i}^{k}-\mathcal{D}_{i}K)=\kappa T_{0i}.
\end{equation}
Inserting (\ref{hypersurefaceij}) we arrive at the momentum constraint

\begin{equation}
\Phi_{i}(\gamma,K)=\frac{\sqrt{\gamma}}{2\kappa}\left(\mathcal{D}_{i}K-\mathcal{D}_{k}K_{i}^{k}\right)-\frac{\sqrt{\gamma}}{N}\left(N^{j}T_{ij}-T_{0i}\right)=0.
\end{equation}
Similarly, the Hamiltonian constraint can be obtained via the $00$
component of the field equations. We write 
\begin{equation}
R_{00}-\frac{1}{2}g_{00}R+\varLambda g_{00}=\kappa T_{00}.
\end{equation}
We insert (\ref{r00}, \ref{hypersurefaceij}) and use $g_{00}=N_{i}N^{i}-N^{2}$
to arrive at 
\begin{equation}
\frac{N^{2}}{2}\left(R-2K_{ij}^{2}-2\Lambda\right)+N(\mathcal{D}_{k}\mathcal{D}^{k}N-\dot{K}-N^{k}\mathcal{D}_{k}K+2N^{K}\mathcal{D}_{i}K_{i}^{k})-\kappa\left(T_{00}+N^{i}N^{j}T_{ij}\right)=0.
\end{equation}
Moreover using (\ref{r}) together with the momentum constraint one
gets the Hamiltonian constraint

\begin{equation}
\Phi_{0}(\gamma,K)=\frac{\sqrt{\gamma}}{2\kappa}\left(-^{\Sigma}R+K_{ij}^{2}-K^{2}+2\Lambda\right)-\frac{\sqrt{\gamma}}{N^{2}}\left(2N^{i}T_{0i}-T_{00}-N^{i}N^{j}T_{ij}\right)=0.
\end{equation}

\end{document}